\documentclass[]{jkas} 

\usepackage[normalem]{ulem}

\def\year{2025} 
\def\volume{56} 
\def\issue{1} 
\def\beginpage{1} 
\def\received{---} 
\def\accepted{---} 
\def\published{---} 
\date{Received \received; Accepted \accepted; Published \published}

\newcommand\halpha{H$\alpha$}
\newcommand\caii{Ca {\sc ii}}
\newcommand\um{$\mu$m}
\newcommand\asc{$^{\prime\prime}$}
\newcommand\kmps{km\,s$^{-1}$}

\title{%
Challan: Solar Full-disk Imaging-Spectroscopic Telescope using the Drift-Scanning Technique
}

\author[1]{Heesu Yang}{0000-0001-5455-2546}
\author[1,2,3]{Maria S. Madjarska}{0000-0001-9806-2485}
\author[1,4]{Donguk Song}{0000-0003-3034-8406}
\author[1,5]{Hannah Kwak}{0000-0001-8619-9345}
\author[1,6]{Sung-Hong Park}{0000-0001-9149-6547}
\author[1]{Eun-Kyung Lim}{0000-0002-7358-9827}
\author[1]{Sujin Kim}{0000-0002-5004-7734}
\author[1]{Su-Chan Bong}{0000-0003-1859-0515}
\author[1]{Yeon-Han Kim}{0000-0001-5900-6237}
\author[1]{Seonghwan Choi}{0000-0002-1946-7327}

\affil[1]{Korea Astronomy and Space Science Institute, 776 Daedeok-daero, Yuseong-gu, Daejeon 34055, Republic of Korea}
\affil[2]{Max Planck Institute for Solar System Research, Justus-von-Liebig-Weg 3, 37077, G\"ottingen, Germany}
\affil[3]{Space Research and Technology Institute, Bulgarian Academy of Sciences, Acad. G. Bonchev Str., Bl. 1, 1113, Sofia, Bulgaria}
\affil[4]{National Astronomical Observatory of Japan, 2-21-1 Osawa, Mitaka, Tokyo 181-8588, Japan}
\affil[5]{Research Institute of Natural Sciences, Chungnam National University, 99 Daehak-ro, Yuseong-gu, Daejeon 34134, Republic of Korea}
\affil[6]{Department of Astronomy and Space Science, University of Science and Technology, Daejeon 34113, Republic of Korea}

\def\corrauthor{%
H. Yang, \email{hsyang@kasi.re.kr}
}

\def\runningauthor{%
Yang et al.
}

\def\runningtitle{%
Challan: Solar full-disk imaging-spectroscopic telescope
}

\def\keywords{%
instrumentation: spectrographs --- Sun: chromosphere --- techniques: imaging spectroscopy
}

\def\abstracttext{%

The Challan instrument is a solar full-disk imaging spectroscopic telescope planned to be installed at three sites with a 120-degree longitudinal difference, enabling continuous 24-hour observations of the Sun. It will take data every 2.5\,min with a spatial resolution of $2-3$\asc\ and a spectral resolving power (R) of $>43,000$ in \halpha~and Ca {\sc ii} 8542\,\AA~bands simultaneously. Challan is composed of two modules, each dedicated to a specific waveband. This modular design is beneficial in minimizing the scattered light and simplifying the structure and engineering. The primary scientific goal of Challan is to investigate solar flares and filament eruptions. It is also expected to detect small-scale events in the solar chromosphere. In 2025,  Challan will be installed at the Big Bear Solar Observatory for test observational runs, followed by scientific runs in 2026.
}

\begin{document}
\jkashead

\section{introduction}
The Sun is the fundamental reference for all stars and is the only star for which spatially resolved spectroscopic observations of various types of activity are available. Ground-based observations of the Sun are primarily focused on the visible light spectrum with images taken in the white light, G~band (4307\,\AA) and TiO bands (7057\,\AA), and in the \halpha~(6563\,\AA), \caii~K and H (3934\,\AA\ and 3969\,\AA), \caii~infrared (IR) triplet (8498\,\AA, 8542\,\AA, and 8662\,\AA), Na~{\sc i} D$_2$ and D$_1$ (5890\,\AA~and 5896\,\AA), He~{\sc i}~D$_3$ (5876\,\AA), Mg~{\sc i} b$_2$ and b$_1$ (5173\,\AA~and 5184\,\AA), He~{\sc i} 10830\,\AA, including various Fe~{\sc i} spectral lines \citep{2000asqu.book.....C}. Among those, we note in particular the \halpha\ and Ca~{\sc ii} IR 8542\,\AA~lines. These lines have been extensively observed using both large and small-aperture telescopes \citep{1964ApOpt...3.1337P,denker_1999,cao_2010,2012ApJ...747..129L,2013RAA....13.1509F,2020SoPh..295..172R}.
The \halpha\ line is a representative line of the chromosphere \citep{1868RSPS...17..128L}. It is one of the most widely used and well-understood chromospheric spectral lines \citep[][]{2012ApJ...749..136L}. The line is the best omnipresent proxy of the shock-rich chromosphere, because the ion population of $n=2$ is large in the shocks and small in cool post-shocked plasma. Also, the excitation rate for the Balmer series shows a strong dependence on the ion population because of the large excitation energy \citep{2008ASPC..397...54R}.

In addition, the Ca {\sc ii} IR line is one of the reliable diagnostic lines of the low chromosphere \citep{2007A&A...461L...1V}. The line has broad wings that sample the photosphere and can be used for photospheric temperature diagnostics \citep{1972SoPh...25..357S,1974SoPh...39...49S}. The line core inscribes solely the chromosphere, showing highly dynamic small bright grains or fibril-like structures when observed at lower resolution \citep{2008A&A...480..515C}.

Large-aperture solar telescopes, with their long focal lengths, are not suited for observing large-scale events such as filaments or flares, because they are designed to capture small and localized regions at high spatial resolution. They also face challenges related to thermal and mechanical stresses, as well as atmospheric seeing.  
Therefore, complex compensation subsystems (e.g., active optics and adaptive optics) must be included.

The Fast Imaging Solar Spectrograph \citep[FISS;][]{chae_2013}, and the Visible Imaging Spectrometer \citep[VIS;][]{cao_2010}, a single Fabry-Pérot etalon filter, installed on the Goode Solar Telescope (GST) at the Big Bear Solar Observatory (BBSO), are used to observe the \halpha, \caii~8542\,\AA, Na~{\sc i} D$_2$ lines or various visible spectral lines, contributing to studies on topics, like for instance, the dynamics of the temperature minimum region and the chromosphere, with aid of the adaptive and active optics. Similarly, the Swedish Solar Telescope (SST) employs dual Fabry-Pérot interferometers to observe the \halpha~and \caii~IR~lines \citep{2008ApJ...689L..69S}. Recently, high-spatial-resolution imaging-spectroscopic observations of lines such as H$\beta$, H$\gamma$, and H$\delta$ have been taken with the SST, allowing studies of small-scale events \citep[for review see][]{2012SSRv..169..181T}, for example, Ellerman bombs \citep{ellerman_1917} or spicules \citep{2012SSRv..169..181T}.

Space-based instruments such as the Chinese \halpha\ Solar Explorer \citep[CHASE,][]{2022SCPMA..6589602L} provide full-disk \halpha\ spectroscopic data. However, their temporal coverage is limited because of constraints on data transmission from space.

Small solar telescopes, dedicated to specific scientific purposes, have been used to study global-scale variations, e.g., solar activity such as flares, filament/prominence eruptions \citep[e.g.,][]{sakurai_1995,malherbe_1983,denker_2007,2018SunGe..13..157S}, or global oscillations of the Sun \citep{harvey_1996}. The observations using small-aperture telescopes have the benefit of acquiring continuous synoptic data for long periods of time \citep{bertello_2020,malherbe_2019} to understand the statistical properties of the small-scale events such as mini-filaments or Ellerman bombs \citep[e.g.,][]{denker_1999}. They also help support space observations \citep{sakurai_1995}, which are mostly dedicated to wavelength bands which are restricted to employ from ground observatories, e.g., the Atmospheric Imaging Assembly \citep[AIA;][]{2012SoPh..275...17L} on board the Solar Dynamics Observatory \citep[SDO;][]{pesnel_2012}, the Interface Region Imaging Spectrograph \citep[IRIS;][]{pontieu_2014}, Soft X-ray Telescope \citep[SXT;][]{tsuneta_1991} on board Yohkoh \citep{acton_1988}, etc.

In this paper, we introduce the Challan instrument dedicated to obtaining full-disk imaging spectroscopic data in \halpha~and Ca {\sc ii} 8542\,\AA~bands by drift-scanning the Sun, which is scheduled for initial installation at BBSO in 2025, with regular operations expected to commence in 2026. The Korean word `Challan' refers to something brilliant, radiant, or glorious, and is often used to describe dazzling light or magnificent beauty. The name reflects the instrument’s ability to capture the Sun’s radiant features. We present the scientific objectives and the system requirements of the instrument in Section~\ref{sec:req}, the concepts of design are described in Section~\ref{sec:concept}, and the assembly and alignment are described in Section~\ref{sec:alignment}. The control software and automation are described in Section~\ref{sec:software}, and data acquisition and processing are introduced in Section~\ref{sec:data}. In Section~\ref{sec:summary}, we summarize the development and prospects of Challan, including the test observations of the \halpha~bands conducted at the Jang Bogo Station.

\section{Scientific Objectives and System Requirements\label{sec:req}}

The primary scientific objective of Challan is to investigate solar flares that affect the solar atmosphere and space weather. When a flare explosively occurs in the solar corona, the non-thermal heating by energetic particles causes plasma evaporation in the chromosphere, generating transient brightenings identified in the Fraunhofer lines \citep{Tian_2018,2016IAUS..320..239C}. By analyzing the variations in the intensity of the spectral lines, we can understand how energetic particles are generated at the reconnection site and how they affect the lower atmosphere by combining observations with predictive models \citep[e.g.,][]{1992ApJ...397L..59C}.

The second objective is to investigate the mechanisms of solar filament eruptions and to explore methods for predicting them. Before eruption, filaments show slow rise \citep[e.g., ][]{2003ApJ...599.1418S,2005ApJ...630.1148S,2006A&A...458..965C,2008ApJ...674..586S}, helical motions \citep{2013ApJ...773..162B}, increase of the internal flows \citep{2017ApJ...843L..24S}, a deformation, or an increase of the amplitude of oscillations \citep{2006A&A...449L..17I,2008ApJ...680.1560P,2014ApJ...790..100B}. A quiescent filament may become unstable due to various factors \citep{2014LRSP...11....1P}, as identified with increasing oscillation amplitude. Above a certain threshold of an increase in amplitude, the filament may then erupt. These oscillations typically last a few hours, slightly decreasing with time \citep{2008ApJ...680.1560P}. Thus, the increase in oscillation amplitude and the decrease in the oscillation period can be used to predict the eruption of a filament. 

An increase in the non-thermal velocity of plasma can also be a precursor to a filament eruption. Such an increase in the non-thermal velocity has been observed before flare onset \citep{2014LRSP...11....1P}. The filament's non-thermal velocity is also expected to increase before its eruption. \citet{2017ApJ...843L..24S,2019PASJ...71...56S} reported that the standard deviation of the small-scale motions in filaments observed in the \halpha\ line increases before the slow rise phase of filaments, which can be used as a proxy of the non-thermal velocity of the plasma. Observations of Challan can be used to estimate the non-thermal velocity by measuring the widths of several chromospheric lines. Since thermal broadening is dependent on ion mass, the non-thermal velocity can be extracted by comparing the widths of spectral lines from ions with different masses \citep{chae_2013}.

The third objective is to understand small-scale energetic events driven by magnetic reconnection and their impact on the upper atmosphere, particularly from a statistical perspective. In particular, we will focus on Ellerman bombs, which have been observed in the Balmer lines and are characterized by a specific line profile exhibiting strong emission in the blue and red line wings \citep{ellerman_1917}. Thus, Challan could provide statistical information on how frequently Ellerman bombs occur, how long they last, and how many are associated with solar chromospheric jets. The differences in their distribution between quiet and active regions \citep{voort_2016}, their correlation with spicules \citep{ola_2025}, as well as their detection across various wavelength bands, are also of special interest. The fine structure of the events will not be resolved due to spatial resolution limitations, however, spectral variations of them can be identifiable. Challan will provide physical parameters with the capacity for observations of various spectral bands.

To achieve these scientific objectives, the instrument should record full-disk spectra at several spectral bands in the visible wavelength range. The spectral resolving power shall be high enough ($>30,000$) to resolve the line shift of the chromospheric lines. To resolve the high-speed plasma motions at the solar \halpha~flares \citep[e.g., up to $130$\,\kmps;][]{2016SoPh..291.2391C} or the solar filament eruptions \citep[e.g., reaching $600$\,\kmps;][]{2024A&A...690A.172C}, the spectral coverage should cover far wings of the spectral lines ($>$20\,\AA). 
The cadence should be a few minutes or less, considering the typical lifetimes of solar flares \citep[$\sim$20\,min;][]{2001A&A...375.1049T} and Ellerman bombs \citep[$\sim$3\,min;][]{2019A&A...626A...4V}. While higher spatial resolution is desirable to resolve the detailed shape of the flares \citep[$0.5-3$\asc;][]{2011ApJ...739...96K,2012PASJ...64...20A} or the area of Ellerman bombs \citep[$\sim$0.15\,arcsec$^2$;][]{2019A&A...626A...4V}, it is ultimately limited by the atmospheric seeing of the observing site, which is expected to be poorer than $3$\asc. Therefore, we compromise the spatial resolution to approximately $3$\asc. The optics and mechanisms should be simple and easy to align, as the instruments will be installed at remote locations. This can be achieved through a modular design for each waveband.

\begin{figure}[ht]
\centering
\includegraphics[width=\linewidth]{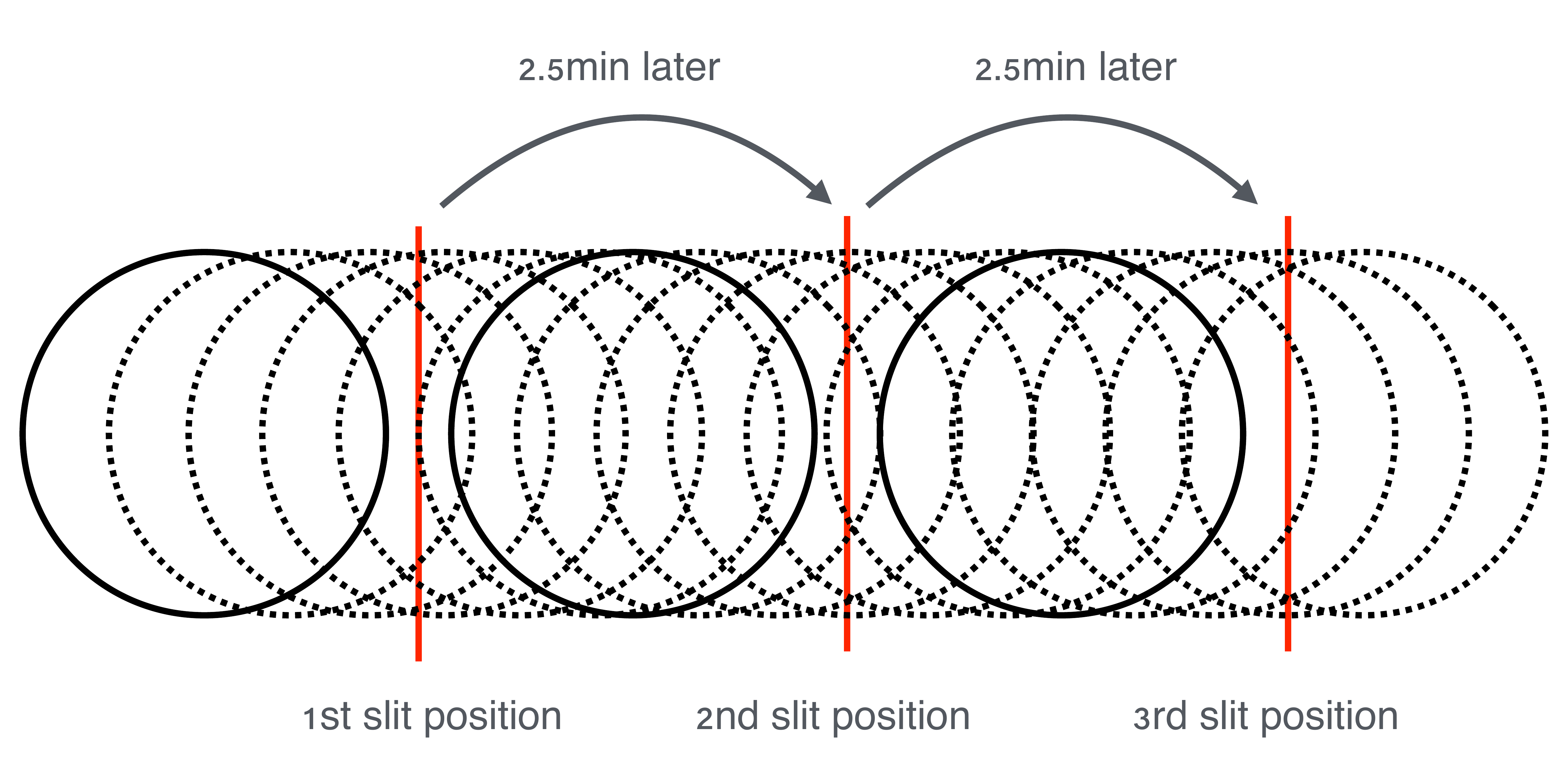}
\caption{Drift-scanning technique. The solid circles represent the solar disks when the slit is positioned at the eastern edge at sequential scanning. The mount is controlled to place the slit at the eastern edge of the solar disk. Then, the mount stops, and the camera captures the solar spectrum for 2.5\,min as the Sun drifts from east to west. }
\label{fig:illust_driftscan}
\end{figure}

\section{Concepts of Design\label{sec:concept}}

\subsection{Technique and Specification}

Challan is designed to obtain solar imaging spectroscopic data using a drift-scanning technique. As illustrated in Figure~\ref{fig:illust_driftscan}, the drift scanning uses the rotational motion of the celestial sphere to scan the solar disk with an accurate scanning speed. A slit of the spectrograph installed on an equatorial mount is positioned along the declination direction at the eastern edge of the solar disk. By halting the equatorial drive and fast capturing the light passing through the slit, we can obtain spectral data across the entire solar disk with an accurate spatial scanning. 

Challan can acquire full-disk data every 150 seconds employing this technique. Each observational cadence includes the sequence of saving the data, moving the slit to the next slit position, and keeping about 8\,\% margin at the east and west limbs of the Sun. Theoretically, the fastest achievable scanning cadence for the full disk using the drift-scanning technique is around 128\,seconds. 

\subsection{Modular Design}

\begin{figure*}
\centering
\includegraphics[width=\linewidth]{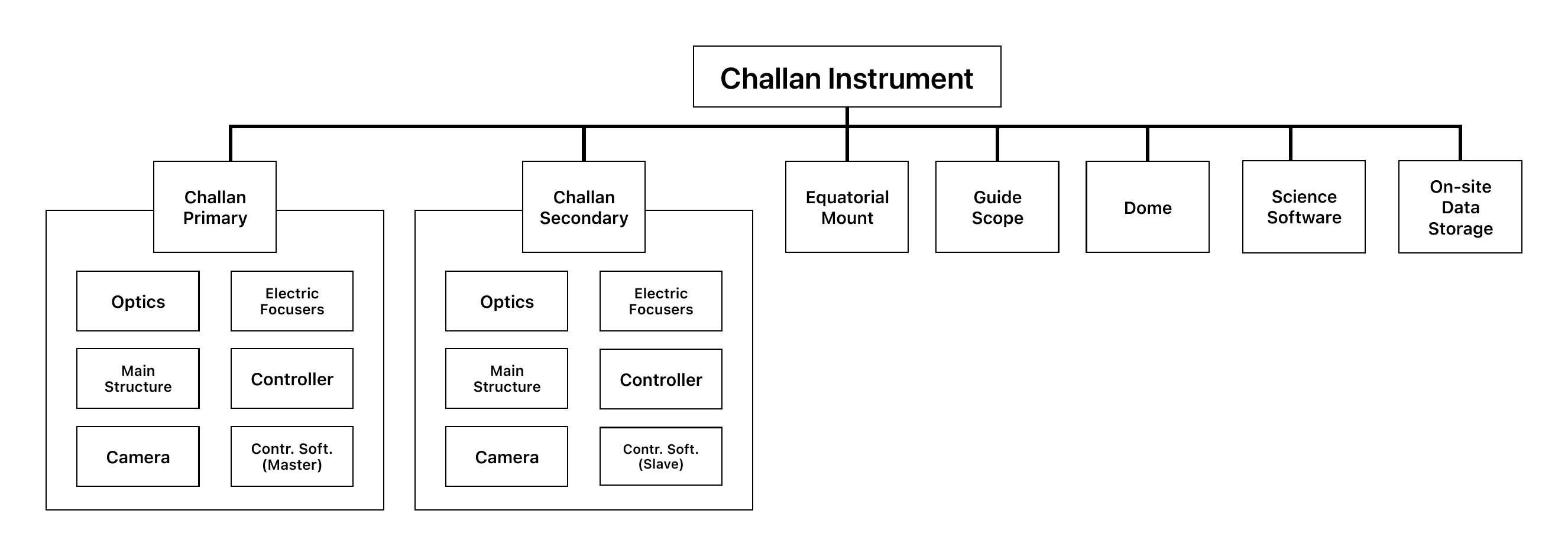}
\caption{Product breakdown structure of Challan.\label{fig:pbs}}
\end{figure*}

Figure~\ref{fig:pbs} shows the Challan's product breakdown structure. The instrument consists of dedicated modules for each observational wavelength. A module includes optics, such as a telescope and a spectrograph, mechanical components, a camera, two electrical focusers, a controller, and control software. To observe several wavebands simultaneously, several modules can be stacked on the same equatorial mount, as shown in Figure~\ref{fig:modular_design}. This modular design is chosen to minimize scattered light and simplify the structure and engineering. We plan to stack two modules targeting \halpha~and Ca {\sc ii} 8542\,\AA~bands, which are well understood by observing the lines using the FISS/GST. Moreover, the data from Challan can provide additional information for the observation of large-scale phenomena using the FISS, such as flares, or filament eruptions.

The controller is composed of a fanless Windows-based computer equipped with a solid-state drive for fast data transfer and storage. Due to the large data size, the speed of internal data handling has to be considered. The controller is connected to the server via a wired network.

The modules operate under a master-slave configuration. The master controller controls the equatorial mount, a guide camera, two electric focusers of the primary module, the spectrograph camera, and the dome. The master controller also triggers the master cameras and is synchronized with the slave camera. Each module passively acquires the data from each spectrograph camera and saves it on its dedicated disk. The slave controller also controls two motorized focusers in the module individually. The control parameters, e.g., the exposure time and the gain of the camera, are transferred from the master controller to the slave controller.

\subsection{Optical Design and Mechanisms}

\begin{figure}[!ht]
\centering
\includegraphics[width=\linewidth]{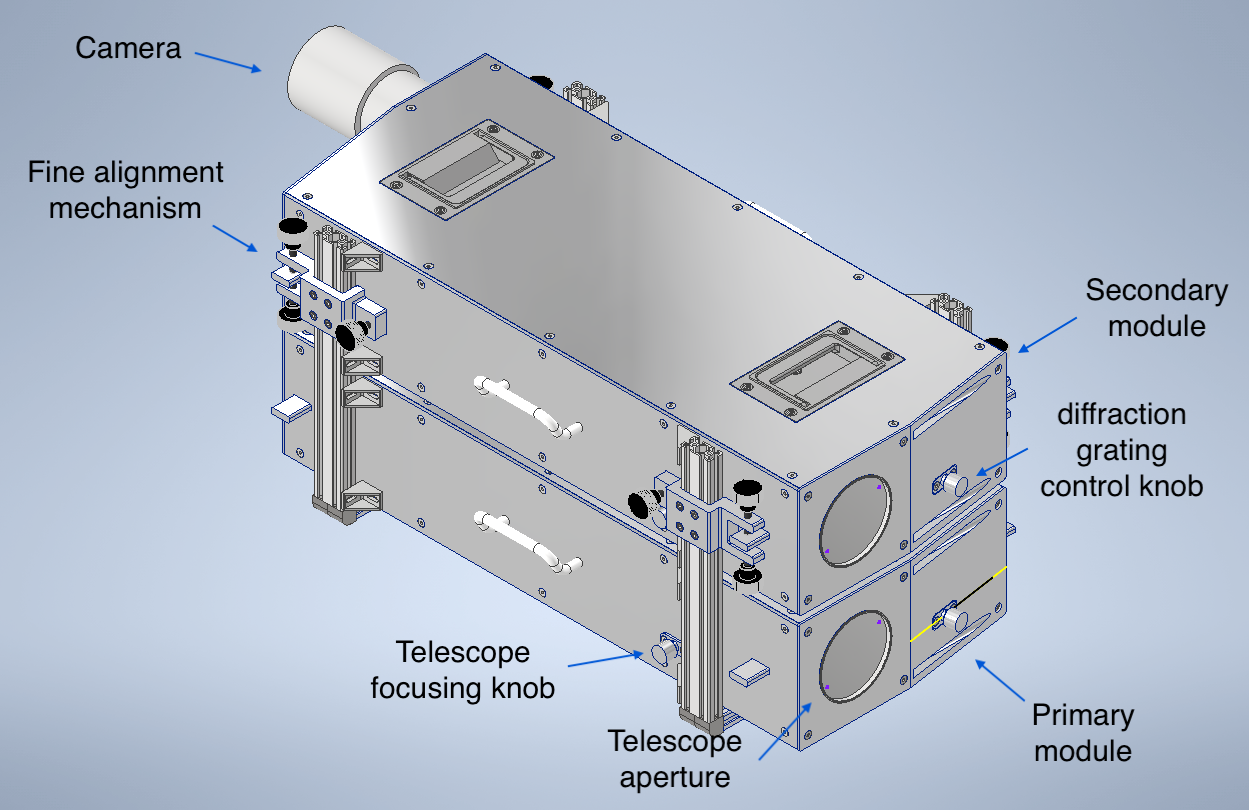}
\caption{3D model of Challan. Two telescope modules are stacked for the multi-band observation.\label{fig:modular_design}}
\end{figure}

\begin{figure}[!ht]
\centering
\includegraphics[width=\linewidth]{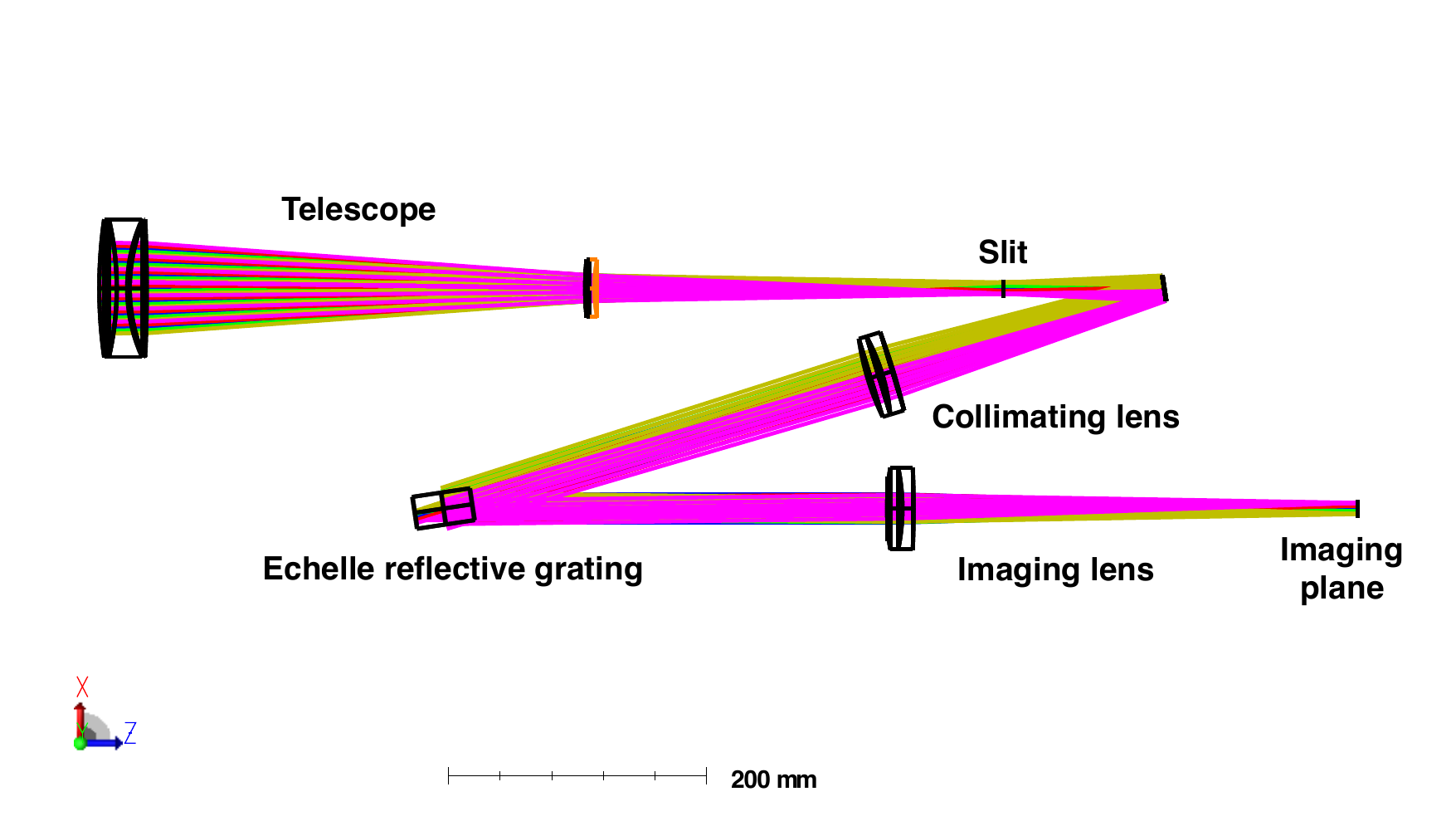}
\caption{Optical layout of Challan. It consists of a telescope and a quasi-Littrow on-axis echelle spectrograph.}
\label{fig:opticallayout}
\end{figure}

Figure \ref{fig:opticallayout} displays the optical design of Challan. It consists of a refractive telescope and an on-axis echelle grating-based spectrograph. All optical parts of Challan are commercial off-the-shelf components. A commercial telescopic camera lens -- {\it Samyang MF-5000 $650 - 1300$\,mm F/8--16 MC IF} -- is utilized for the forefront solar-imaging telescope of Challan. This lens allows for adjustment over a wide range of focal lengths and is easy to focus at the fixed focal plane position. Although this approach makes it challenging to estimate the optical performance accurately, we expect that the reasonable manufacturing tolerance of the camera lens provides acceptable performance regarding various aberrations.

We choose the Littrow configuration, where the incident beam is collimated onto a diffraction grating, and the diffraction beam is then fed back into the imager. In this configuration, the angle between the incidence beam and the angle of the diffraction beam becomes closer to the blazed angle. The slit of the spectrograph is $12$\,mm in length and $10$\,\um~in width. The light passing through the slit is directed to a collimating lens via a folding mirror. The collimating and imaging lenses are achromatic lenses with a focal length of $354$\,mm. The same focal length of the lenses is a reasonable solution to minimize the aberrations. A diffraction grating is positioned at the exit pupil to minimize degradation by the surface quality of the grating on the image plane.

We used the echelle reflective grating of 79\,grooves/mm and 63\,degree blaze angle, manufactured by {\it Edmund Optics}. The grating is mounted on the rotating stage to adjust the observing waveband. The rotating stage is controlled manually using a knob installed at the front of the module (see Figure \ref{fig:modular_design}). One unique feature of Challan is the use of an on-axis optical system, where the light is redirected by a diffraction grating perpendicular to the dispersion direction. This approach helps reduce the alignment difficulties and the aberrations caused by the off-axis system. However, this also results in diagonal dispersion of the slit spectrum and weak astigmatism, adding complexity to post-processing and degrading the spectral and spatial resolution.

Due to the high dispersion characteristics of the echelle diffraction grating, we installed a bandpass filter in front of the camera to prevent contamination by the other order of light:  for instance, we installed a filter centering at a wavelength of 6563\,\AA~and with a full-width-half-maximum (FWHM) of 30\,\AA~manufactured by {\it Andover, CO.} for the \halpha\ band.

We selected the scientific cameras, {\it Apollo-M MAX Mono} camera developed by {\it Player One Astronomy}, which features a {\it Sony IMX432} monochrome Complementary Metal-Oxide-Semiconductor (CMOS) sensor with global shutter. The pixel size of the sensor is $9$\,\um $\times9$ \,\um~accommodating a well depth of $100,000$\,electrons with the resolution of $1608\times1104$ pixels. For the observations, we saved only the selected region of interest to reduce data size and increase the acquisition rate.

\begin{table}[t!]
\caption{\label{tab:calculation}Theoretical calculations of spectral resolution of the Challan.}
\centering
\begin{tabular}{lccc}
\toprule
Parameters & \halpha~band & \caii~band \\
\midrule
Wavelength ($\lambda$) & $6563$\,\AA & $8542$\,\AA \\
Order & $34$ & $26$ \\
Pixel size & $43$\,m\AA&$56$\,m\AA\\
Grating resolution ($\delta\lambda_g$) & $4.9$\,m\AA &$83$\,m\AA\\
Spectral purity ($\delta\lambda_s$) & $119$\,m\AA& $156$\,m\AA\\
Spectral resolution ($\delta\lambda$)&$129$\,m\AA& $177$\,m\AA \\
Resolving power ($R$)& $50,874$&$48,260$ \\
\bottomrule
\end{tabular}

\end{table}

Table \ref{tab:calculation} shows the theoretical calculations of the spectral performance of the instrument. The net spectral resolution can be given as the root mean square of the grating resolution and the spectral purity.
In our system, the spectral purity, which is defined by the smoothening from the entrance slit width of $10$\,\um, dominates the spectral performance. As a result, the instrument theoretically can achieve the resolving power of $\sim50,000$.

The focus of the telescope is optimized manually using a knob installed on the side of the module. The knob rotates the focusing ring, located in the middle of the telescope, without changing the position of the focusing plane. Using the micro-focuser installed in front of the camera, the focus of the spectrograph is manually optimized. For the remote observations, we install electric focusers at the rotating knobs for the focus of the telescope and the spectrograph.

Challan utilizes an equatorial mount, {\it RST-300}, manufactured by {\it Rainbow Astro}, for the drift scanning. The mount operates smoothly and precisely, pointing to the Sun with a strain wave gear and a harmonic drive. 

\begin{table}[t!]
\caption{Specification of the Challan.}
\centering
\begin{tabular}{ll}
\toprule
Parameters & Value \\
\midrule
Field of view & Full-disk \\
Effective aperture & $82$~mm \\
Spectral band & \halpha\ and  \caii~IR (switchable)\\
Spatial resolution  & $\sim2$\asc\ (scan direction)\\ 
                    &$\sim3$\asc\ (slit direction) \\
Wavelength coverage & $>20$\,\AA\\
Cadence      & $\sim2.5$\ min\\
Frame rate & 15 fps \\
Exposure time  & $\sim10$\ ms \\
Data dimension & $1608\times 2232\times552$ pixels\\
Physical pixel size & $9$\,\um $\times$ $9$\,\um\\
Data size                  & $3.7$\,gigabyte\\
\bottomrule
\end{tabular}

\label{tab:specification}
\end{table}

Table~\ref{tab:specification} summarizes the specifications of Challan. The effective aperture of Challan is $82$~mm, combined with an F/11 telescopic lens. The spatial resolution is $2$\asc~for the scanning direction and $3$\asc~for the slit direction. By maintaining a consistent frame rate for the camera, we can achieve the correct spatial resolution in the scanning direction. The spatial resolution of the slit direction depends on the diffraction limit of the telescopic lens and the pixel sampling. 

\section{Optical Alignment\label{sec:alignment}}

Initially, the system is precisely aligned in the laboratory using a laser. However, the system requires a precise on-site alignment after delivery to the observing site or after a long period of operation. Therefore, we developed the alignment mechanism that can be operated at the observing site using sunlight. 
 
The optical alignment starts from the camera side. The distance between the imaging lens and the camera can be adjusted by the incident parallel beam in front of the imaging lens. We temporarily install a folding mirror in front of the imaging lens to direct sunlight to the camera through the imaging lens. Then, we can adjust the position of the imaging lens by making a clear solar image with the camera. After that, we point the telescope at the Sun. The distance between the slit and the collimating lens is adjusted by making a sharp solar spectrum. We note that the distance between the diffraction grating and the lenses is insensitive to the quality of the spectrum. The installation angles of all lenses can be controlled within mechanical tolerances because the on-axis optical system of Challan is beneficial in mitigating axial misalignment. 

All telescope modules for each waveband are mounted on the same equatorial mount to facilitate multi-wavelength observations. The primary module is located at the base, and the secondary module is stacked at the top of the primary module. They are mounted to a frame that encloses both the primary and secondary modules. The primary module is fixed to the frame with screws at each side. The secondary module is aligned with the primary module using specially designed fine alignment components, mounted on four edges of the module, shown in Figure \ref{fig:modular_design}. Once aligned, the position is secured with locking screws at each end. 

\section{Control Software \& Automation\label{sec:software}}


Challan is controlled through a Flask-based web application that serves as a control and monitoring dashboard. The master controller for the primary module controls the telescope mount, the guide camera, the science camera of the module, and the motorized focusers. The cameras are controlled through the drivers provided by the manufacturer, and the other components -- the mount and the focusers -- are controlled through the Windows-based {\it ASCOM} driver. 

The slave controller also generates its web server using the Flask application. On this webpage, observers can manually control the science camera and motorized focusers. The webpage also provides the real-time status of the devices. The slave controller web page is linked to the primary web page, so that one can access every controller through the web pages of the primary controller. 

\begin{figure*}[!th]
\centering
\includegraphics[width=\linewidth]{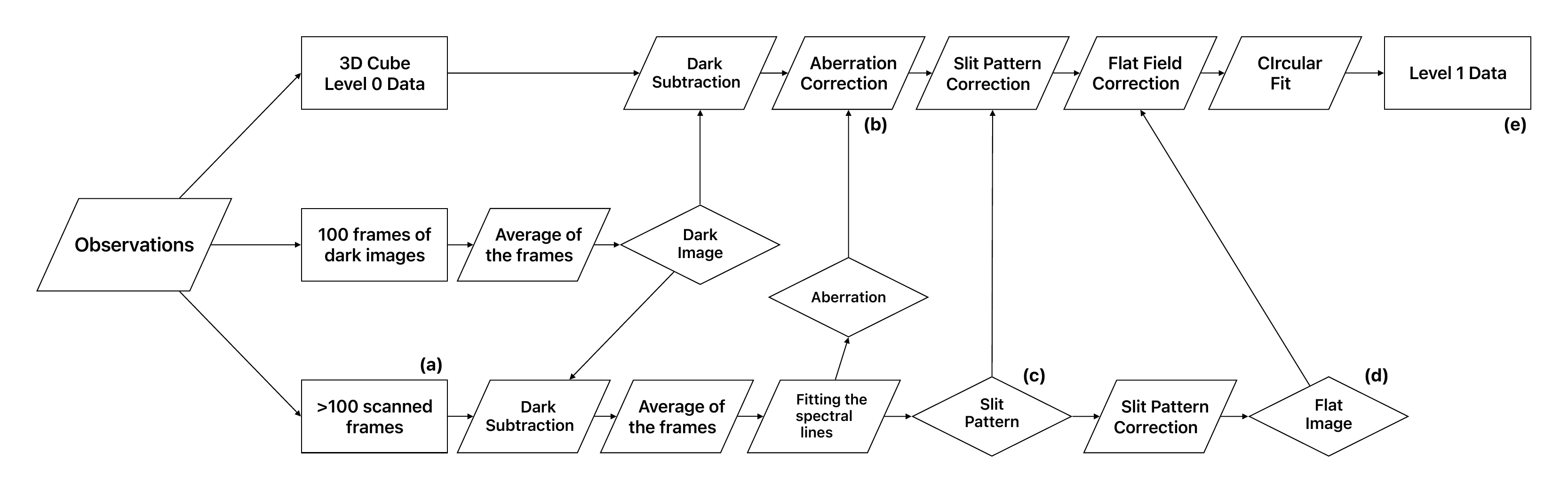}
\caption{Block diagram of the data processing after observations. The labels from (a) to (e) mark the process of the images shown in Figure~\ref{fig:prc_img}. \label{fig:block_diagram}}
\end{figure*}

The observations are fully automated, including opening the dome, focusing the telescopes and spectrographs, and observing the Sun. Before the daily observations, data for post-processing, such as flat-field and dark current, are also automatically collected. 

Once automate observation is started, the equatorial mount positions the slit at the east side of the Sun. Then, the science camera takes images with an exact frame rate. After scanning an area of 0.62 degree for 2.5\,min using the solar rotation, the equatorial mount repositions the slit the east side of the Sun again.

\section{Data Acquisition and Processing\label{sec:data}}

Figure~\ref{fig:block_diagram} shows a block diagram of the data acquisition and post-processing. The dark image is produced by averaging 100 images taken without light. Because Challan does not have a camera shutter or an aperture door, we take dark images before opening the dome. Although the camera does not have temperature stabilization mechanisms, the variation of the dark current is negligible during the observations due to the short exposure time of less than 10\,ms.

\begin{figure}[htbp]
\centering
\includegraphics[width=\linewidth]{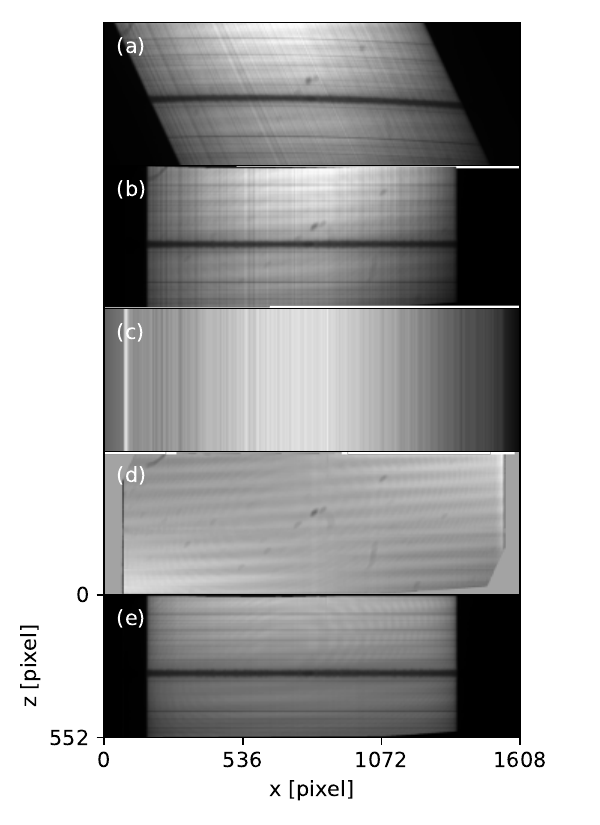}
\caption{(a) Averaged image of 100 spectrum images used to obtain the flat field, (b) Aberration corrected image, (c) Slit pattern, (d) Flat field, and (e) Final corrected image.\label{fig:prc_img}}
\end{figure}

The flat field is obtained using more than 100 spectrum images, scanning the Sun in the north-south direction. After subtracting the dark image from each image, we produce an image that is the average of all 100 images. Figure~\ref{fig:prc_img}(a) presents the averaged image. The absorption spectral lines of the averaged image show a diagonal pattern caused by the reflection vertical to the dispersive direction of the diffraction grating. The absorption lines are also curved, resulting from aberrations. The spectrum, however, has the same spectral shape over the overall wavebands in the averaged image because the spectral variations of the solar atmosphere are smeared out when the spectrum images are averaged. Then, we extract and correct the spherical aberration and image distortion, by fitting the curveture of the absorption lines along the slit direction. Figure~\ref{fig:prc_img}(b) shows the aberration-corrected image. Then, the slit pattern is extracted by averaging the intensity along the dispersion direction. The slit pattern is expanded in the slit direction to obtain the same dimension as the original frame displayed in Figure~\ref{fig:prc_img}(c). Then the slit pattern is divided by the aberration-corrected image. The only remaining pattern is the averaged solar spectrum. Finally, the averaged spectrum is subtracted from the spectra at each slit position, obtaining the flat-field image as shown in Figure~\ref{fig:prc_img}(d). We note that the vignetting caused by the optical components is also corrected during the flat-field correction.

The science level 0 data acquired during the observations are saved as unsigned $16$\,bits FITS files. Every frame of the data is corrected for the dark current, the aberrations, the slit pattern, and the flat-field. Figure \ref{fig:prc_img}(e) displays a sample of the corrected spectrum images. The typical dimension of the data is $1608\times2232\times552$ pixels, where $1608$ pixels represent the dimension along the slit, $2232$ pixels represent the number of steps scanned in 2.5\,min. The $ 552$-pixel dimension corresponds to the spectrum dispersion. During the observations, we save only $552$ pixels, which is half of the original sensor size of $1104$ pixels, thereby optimizing the data transfer and saving time.

The final step is to reconstruct the solar circular disk. The images at specific wavelengths do not show a circular Sun due to the misalignment of the slit to the declination axis and the tilting of the light by the diffraction grating. First, we determine the disk limb using a Sobel filter. Then, we iteratively apply an ellipse fitting to the limb pixels using a least-square approximation to extract: the pixel scales of x- and y-directions, the center of the solar disk, the rotation angle of the ellipse, and the lengths of semi-major and semi-minor axes. Finally, the data are shifted in the x- and y-directions.

We note that the final spectrum image in Figure~\ref{fig:prc_img}(e) shows a fringe pattern superimposed on the processed spectrum image. The flat field does not subtract the pattern because it changes depending on the variations of the optics due to the weight balance during the observations. The pattern is caused by the internal reflection inside the pre-filter located at the front of the camera. It can be fully removed by tilting the pre-filter away from the normal incidence. The fringe pattern can also be removed by Fourier filtering or wavelet-based filtering techniques, as it exhibits concentric rings \citep{1995ASPC...81..138M,2025JKAS...58...63K}.

\begin{figure*}[ht]
\centering
\includegraphics[width=\linewidth]{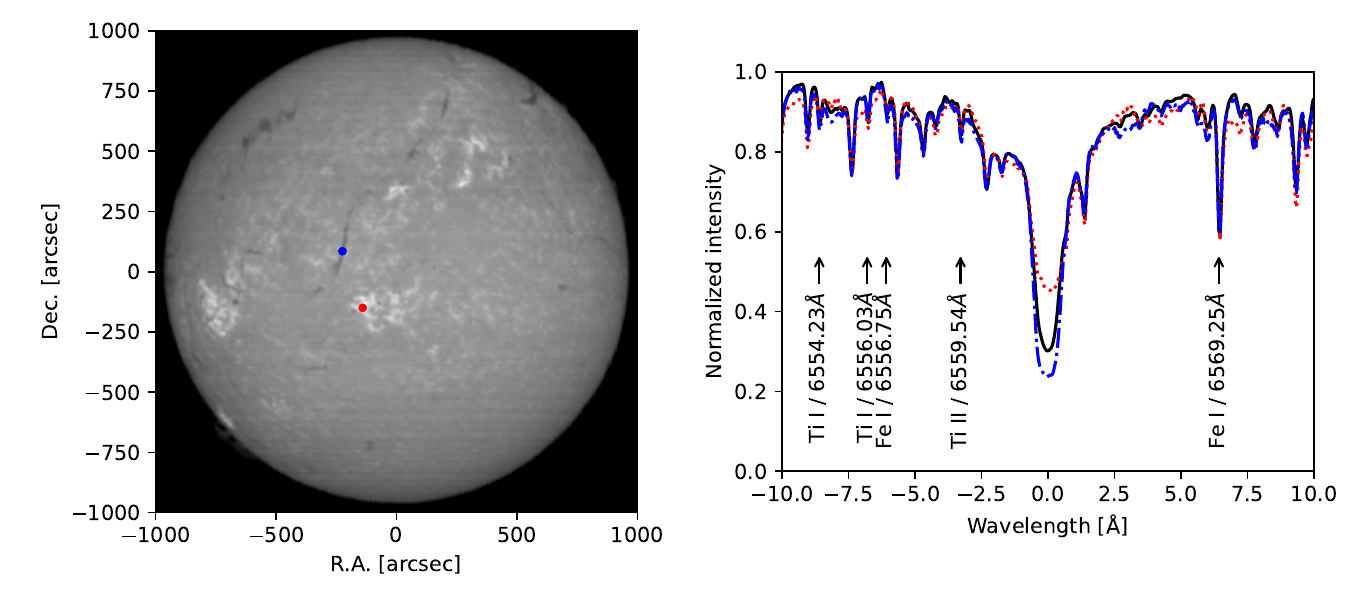}
\caption{\label{fig:fulldisk_img} (a) Reconstructed image at the central wavelength of the \halpha~after post-processing. (b) Corresponding spectra over a 20~\AA\ range centered at the  6563\,\AA$\pm10$\,\AA\ for three regions: the center of the solar disk (black solid line), a filament (blue dash-dotted line), and a plage region (red dotted line). The filament and plage locations are marked in panel (a) with blue and red dots, respectively. The observations were conducted at the Korea Astronomy and Space Science Institute campus in Daejeon, South Korea.}
\end{figure*}

Figure \ref{fig:fulldisk_img}(a) displays the final reconstructed image at the central wavelength of the \halpha\ band. The spatial resolution in the slit direction is about 3\asc, which is double the pixel sampling of 1.5\asc. The spatial resolution in the scan direction is 2\asc, which depends on the scanning frame rate of 15\,frame/second, which corresponds to the sampling in every 1\asc. However, the spatial resolutions in both directions mainly depend on the seeing condition of the observing site, which is usually larger than the instrumental performance. 

Figure \ref{fig:fulldisk_img}(b) shows the spectra of the Sun center, a filament, and a plage region. Their respective spectra show that the depth of the core of the \halpha~line changes with the observed phenomenon. The spectra also show various solar photospheric lines, e.g., Fe~{\sc i} (6556.75\,\AA, 6569.25\,\AA), Ti~{\sc i} (6554.23\,\AA, 6556.03\,\AA), Ti~{\sc ii} (6559.58\,\AA), and strong terrestrial lines. Using these lines, we estimated the spectral resolving power of the instrument. The average FWHM of the lines is estimated at $0.15$\,\AA, which corresponds to the resolving power of $43,000$. We note that the resolving power of the instrument is better than this value because the width of the absorption lines is the combination of the width of the line itself and the degradation of the instrument. 

\section{Summary and Conclusion\label{sec:summary}}

The Challan instrument is a solar imaging spectroscopic telescope that employs the drift scanning technique to scan the solar disk every 2.5 minutes. Its scientific objectives are to investigate solar flares, filament eruptions, and small-scale events in the chromosphere. Challan is designed to observe the \halpha~and Ca {\sc ii} 8542\,\AA~wavebands. The instrument comprises modularized subsystems, each dedicated to a specific waveband. It is designed to operate remotely and automatically. 

The Challan’s most fascinating performance is a spectral resolution of $>43,000$, which provides spectral information of plasma dynamics associated with solar flares and filament eruptions. Its spatial resolution is 2--3\asc, that is compromised with the seeing conditions at the observing site.

To achieve its scientific objectives, continuous monitoring of the Sun is required. We plan to deploy Challan at Mount Bohyun in South Korea, the Big Bear Solar Observatory in California, and a site in Europe. The installation of three instruments with a 120-degree longitudinal difference will enable continuous 24-hour imaging spectroscopic observations, a capability not previously achieved.

One of the most significant challenges associated with 24-hour operation is the generation of substantial data size. A single data set comprising $1608\times2232$ pixels, coupled with spectral information for $552$\,pixels, results in the data size of $3.7$\,gigabyte. If data are generated every $2.5$\,min in two wavelength bands, this results in $2.1$\, terabytes for an 8-hour observation per day. Additionally, substantial computational resources are necessary to process the data. A potential solution for dealing with the massive data is transferring the data to South Korea after the observations. The data can then be processed using the graphics processing unit (GPU).

\begin{figure}
\centering
\includegraphics[width=\linewidth]{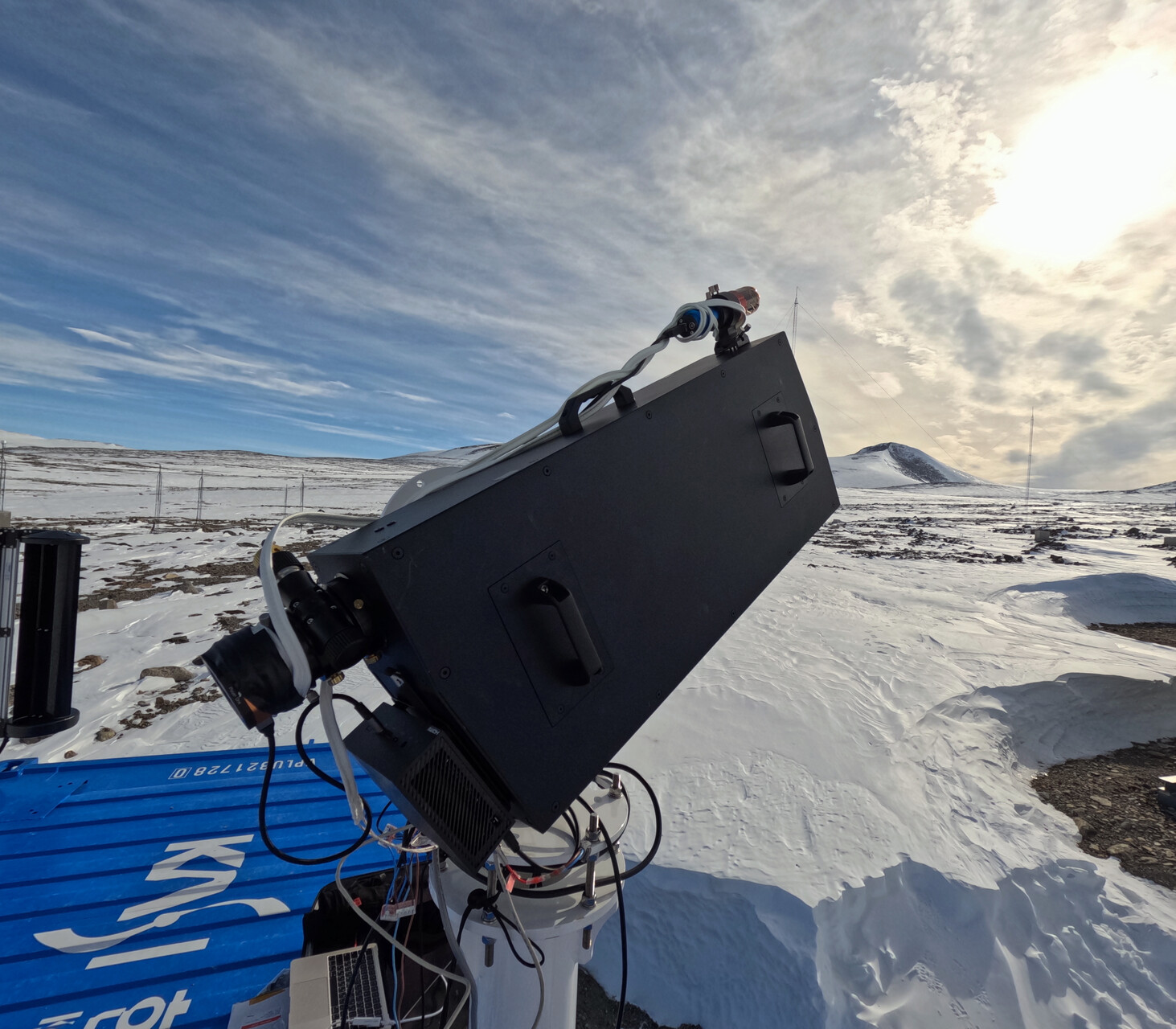}
\caption{Challan installed at Jang Bogo station in Antarctica. The observing campaign was done with a single module dedicated to \halpha~bands. \label{fig:telescope_antarctica}}
\end{figure}

The initial remote observation test was successfully conducted from the inland space weather monitoring module, which was temporarily installed next to the space environment observation building at Jang Bogo Station in Antarctica, as shown in Figure~\ref{fig:telescope_antarctica}. The instrument was placed at an elevation of approximately 3 meters above the ground. Due to logistical constraints associated with the substantial size of the instrument, only a single module using the \halpha\ band was tested. In 2025, we anticipate installing a Challan at the BBSO.  After the installation, we will carry out test observation runs for several months. These runs will be used to optimize and debug the instrument. After the test is completed, the science run will start in 2026.  

\acknowledgments

This work was supported by the National Research Foundation of Korea(NRF) grant funded by the Korea government(MSIT) (RS-2022-NR071796). M.M. acknowledges the support of the Brain Pool program funded by the Ministry of Science and ICT through the National Research Foundation of Korea (RS-2024-00408396) and  DFG grant WI 3211/8-2, project number 452856778. H.K. was supported by Basic Science Research Program through the National Research Foundation of Korea(NRF) funded by the Ministry of Education(RS-2024-00452856).




 \bibliography{ms_hsyang}






\end{document}